\documentclass[submission,copyright,creativecommons]{eptcs}

\usepackage{underscore}           

\usepackage{amsmath}
\usepackage[autostyle]{csquotes}
\usepackage{textcomp} %
\usepackage{prftree}
\usepackage{amssymb} %
\usepackage{mathpartir}
\usepackage{rotating} %
\usepackage{graphicx} 
\graphicspath{ {imgs/} } %
\usepackage{subcaption}
\usepackage[disable]{todonotes}
\usepackage{cleveref}
\usepackage{amsthm}
\theoremstyle{definition}
\crefname{mylist}{list}{lists} %
\theoremstyle{definition}
\newtheorem{definition}{Definition}[subsection]

\title{A logical framework with a graph meta-language} %
\author{Bruno Cuconato\textsuperscript{1,}\thanks{The author thanks CNPQ for its financial support.} \qquad Jefferson de Barros Santos\textsuperscript{1,2} \qquad Edward Hermann Haeusler\textsuperscript{1} %
\institute{\textsuperscript{1} Departamento de Informática, PUC-Rio\\ Rio de Janeiro, Brazil}
\institute{\textsuperscript{2} Escola Brasileira de Administração Pública e de Empresas, FGV-RJ\\
Rio de Janeiro, Brazil %
\email{\quad \{bclaro,jsantos,hermann\}@inf.puc-rio.br}}
}

\newcommand{\myref}[1]{\cref{#1}} %
\newcommand{\mynote}[1]{\todo[inline,color=green!40]{#1}}
\newcommand{\jeffi}[1]{\todo[inline,color=blue!40]{#1}}

\newcommand{\glf}{GLF} %

\begin{document} %
\maketitle

\begin{abstract}
  We conjecture that the relative unpopularity of logical frameworks among practitioners is partly due to their complex meta-languages, which often demand both programming skills and theoretical knowledge of the meta-language in question for them to be fruitfully used. %
  We present ongoing work on a logical framework with a meta-language based on graphs. %
  A simpler meta-language leads to a shallower embedding of the object language, but hopefully leads to easier implementation and usage. %
  A graph-based logical framework also opens up interesting possibilities in time and space performance by using heavily optimized graph databases as backends and by proof compression algorithms geared towards graphs. %
  Deductive systems can be specified using simple domain-specific languages built on top of the graph database's query language. There is support for interactive (through a web-based interface) and semiautomatic (following user-defined tactics specified by a domain-specific language) proof modes. %
We have so far implemented nine systems for propositional logic, including Fitch-style and backward-directed Natural Deduction systems for intuitionistic and classical logic (with the classical systems reusing the rules of the intuitionistic ones), and a Hilbert-style system for the K modal logic. %
\end{abstract}

\section{Motivation and introduction}
\label{sec:motivation}


Deductive systems are regularly proposed without an accompanying implementation (\cite{brock-nannestad2014hybrid,olarte2020fresh,pientka2019type,veloso2013graph,stolze2017towards} are examples). %
After a deductive system is developed, sometimes an implementation is made, almost as an afterthought. %
We believe, however, that having such implementations available during the development of such a system is valuable, and even after the system has been fully developed, an implementation is still helpful in understanding it --- not to mention in applying it. %
We postulate that this problem is pervasive because the cost of developing a custom deductive system (or adapting an existing one) is high. %

Deductive systems are defined by their axioms and rules of inference. %
Logical frameworks (LFs) are meta-languages used to specify deductive systems, as defined by \cite{pfenning1996practice}. %
Logical frameworks allow us to carry out derivations in the implemented deductive system, to implement algorithms that solve a problem related to the implemented deductive system, and some frameworks\todo{Not ours, though…} even allow the user to investigate the meta-theory of the implemented systems (e.g., proving that a system possesses a particular property). %
Besides these advantages, logical frameworks reduce the cost of implementing a deductive system by supplying a common base upon which all systems can build. %
There's another advantage to having a shared codebase besides reducing implementation costs: since logical frameworks are critical software where bugs are unacceptable (unlike, say, a social media app), it is common practice to have a small kernel that is easily verifiable by the third parties and is where the main logic of the application code lies. %
The more deductive systems implemented using the same kernel, the more robust the LF's kernel becomes, which benefits all the superjacent systems. %
Despite the existence of several logical frameworks with different meta-languages, their use by practitioners (e.g., mathematicians) is still rare. %
\cite{buzzard2020perfectoid} seems to be the first formalization of contemporary advanced mathematics (in that it involves a sophisticated object only taught at graduate-level) in a logical framework; most other achievements were long proofs about more elementary objects (e.g., the proof of the four-color theorem \cite{gonthier2007colour}), or formalizations of theorems about more elementary mathematical objects (e.g., \cite{gouezel2019corrected}'s formalization of the Morse lemma). %
To the best of our knowledge, no investigation of why logical frameworks have not been widely adopted so far has been conducted. 

\mynote{shortcomings of most (all?) LFs} %
The present work is no such investigation; however, it is difficult to imagine LFs in widespread use before their shortcomings are surmounted.%
\footnote{Surmounting these problems are no guarantee of success, of course.} 
Most LFs have a user interface problem, having been designed by computer scientists but having a larger intended audience. %
Some of the most popular LFs are programming language environments, which may impose an entry barrier to some (and even experienced programmers sometimes have trouble with the kind of programming languages such LFs employ). %
The need to learn the LF's meta-language, especially when it is a sophisticated one, also adds to this barrier. %
Another problem is that formalizing a pen-and-paper system or proof is rarely a straightforward matter. %
It is not that a lot of implicit details need to be filled in (this is expected although it might be surmountable), but that the subject of the formalization often needs to be slightly (or even not so slightly) adjusted to suit better the LF in question (this point is briefly brought up by \cite[§10]{buzzard2020perfectoid}). %

These shortcomings in logical frameworks point to possible improvements, some of which we try to address on our logical framework (named \glf{}). %
\glf{}'s metalanguage (see \cref{sec:meta-language}) is based on graphs (as is \cite{schroeder2008graph}'s LF), for their simplicity and expressivity. %
Deductive rules are graph transformations, and proofs are represented as graphs in a graph database backend (see \cref{sec:proof-representation}). %
Its user interfaces (see \cref{sec:ui}) attempt to cater to non-programmers, even if programmatic interfaces are also available. %
We strive to have deductive systems implemented in \glf{} be close to their pen-and-paper versions (at the cost of more difficult formal verification of their correctness), but also to offer semi-automatic capabilities with the definition of proof tactics (see \cref{sec:expressivity}). %
We have not been able yet to mitigate all the shortcomings in LFs that we would like to; \cref{sec:future} discusses some of the work that is still to be done in this respect. %


 %


\section{The graph logical framework overview}
\label{sec:graph-lf} %

\begin{figure}
  \centering %
  \makebox[0pt]{\includegraphics[width=1.2\linewidth]{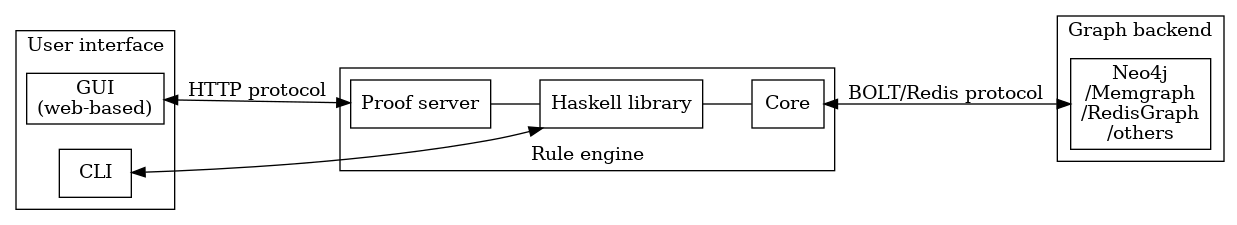}}
  \caption{Architecture overview of the graph logical framework}
  \label{fig:architecture}
\end{figure}

The logical framework herein described is built as a Haskell library implementing the rule application engine, supported by a graph database backend. %
A choice of graph databases is available, provided they implement either of the open standard protocols Bolt or Redis. %
User interfaces are separate from the core LF, and are described in \cref{sec:ui}. %
An overview of the system architecture of \glf{} can be seen in \myref{fig:architecture}. %


Deductive systems in \glf{} are sets of rules with some metadata like name and example formulas, a set of proof strategies for the system, and a function specifying how to parse logic formulas in that system. %
Formula parsers are S-expression-based and derived automatically from a specification of the operators of the system, or else user-defined. %
Deductive rules in \glf{} are a graph database query, a specification of the rules arguments (which may be a graph database query that selects an argument or collects possible arguments in the case of rules with enumerable arguments), plus some metadata like the rule's name. %
The graph database language is the base of the meta-language used by \glf{}, which is described in \cref{sec:meta-language}. %
Currently, deductive systems are specified at compile-time using declarative Haskell code (i.e., something akin to Javascript object language), but dynamically declaring systems (e.g., from configuration files) is possible and remains to be implemented. %

\subsection{User interfaces}
\label{sec:ui} %
Following \cite{TBK92}'s advice, we separate the `core' of the LF from its user interfaces. %
The core itself offers only one programmatic interface: its use as a Haskell library by other Haskell programs. %
On top of this interface, we build an API served over HTTP that can be accessed from any programming language and a somewhat limited command-line user interface (CLI). %
The HTTP API is the backend of a web-based graphical user interface implemented in Javascript. %
This web-based interface we expect most end-users to employ, while the CLI is mainly used for administrative (e.g., starting the proof server) and testing purposes. %

The web-based \glf{} interface is a multi-user web application served over a secure HTTPS connection with user authentication.\footnote{An instance is available at \url{https://glf.tecmf.inf.puc-rio.br/}, while the source code is at \url{https://gitlab.com/odanoburu/glf}.} %
Its current state is functional, but no attention has been paid to styling it or keeping its user latency low. 
Non-concomitant user collaboration is possible, but features like real-time collaboration have not been implemented. 
The architecture is similar to that of \cite{Kaliszyk2007webproof,santos2010infraestrutura}, with a more modern technology stack. %
Interface-wise it is very different from \cite{Kaliszyk2007webproof}, primarily due to the nature of the metalanguages of \glf{} and Coq (\cite{coq2017}). 

The actual \glf{} deduction interface is multifaceted: for each `kind' of deductive system, we strive to offer an interface that closely resembles its pen-and-paper version. %
There are currently two such kinds of deductive systems with web-based interfaces: systems based on backwards-directed (from goal to hypotheses) Natural Deduction respecting the sub-formula principle (as in \cite[§2, def. 1-4]{haeusler14subformula}), and forward-directed Fitch-style deductive systems. %

\begin{figure}
  \centering
  \includegraphics[width=0.9\linewidth]{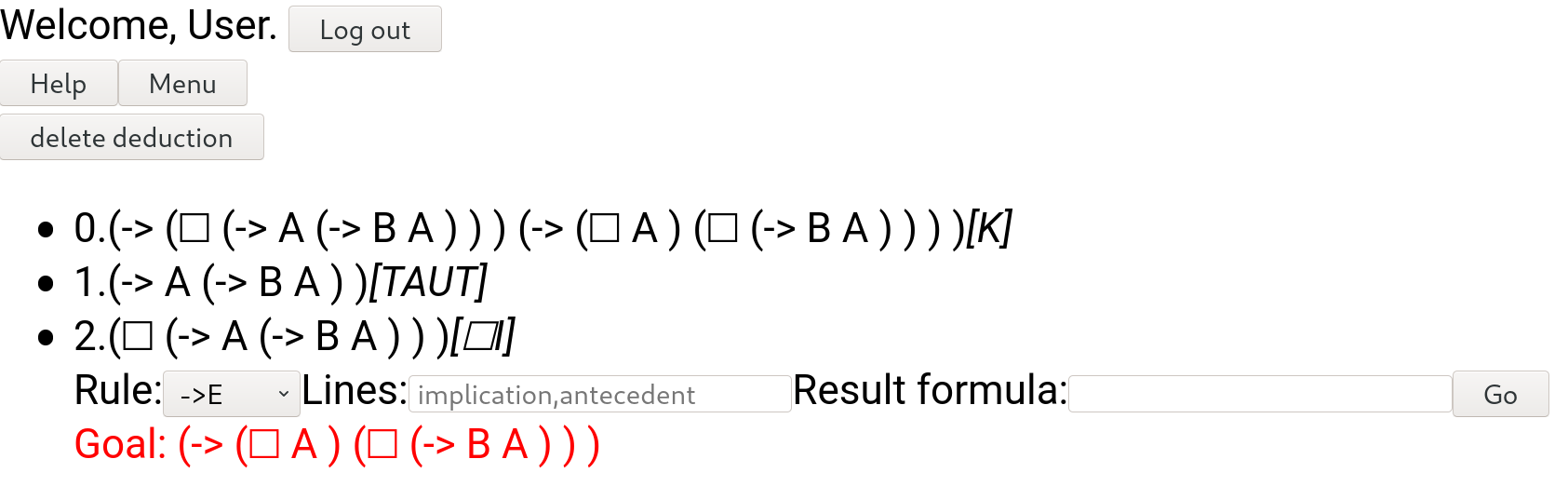}
  \caption{Web-based interface for a Fitch-style K modal system}
  \label{fig:gui-fitch-example}
\end{figure}

Both the Fitch and the Natural Deduction interfaces heavily rely on the mouse, but not exactly as in `proof by pointing' (as defined in \cite{BKT94}). %
Proofs are tree-shaped (as expected), and each node is labeled by the formula it derives (or whose goal it is to derive). %
In the case of Fitch-style systems, all interaction is centered on a single form where the user may select the rule to be applied; depending on the choice, arguments may have to be chosen from a rule-specified enumeration, or the proof lines to which the rule is to be applied must be inputted; finally, the resulting formula may need to be inputted, if it cannot be guessed from the previous information (see \cref{fig:gui-fitch-example}). %
For backward-directed Natural Deduction systems, nodes are color-coded according to their status (goal, leaf, regular), and there is one interaction area for each goal node, from where a user may select a rule and arguments to apply to the goal at hand.%
\todo{add link to our instance, with login details for review?} %



\subsection{Meta-language}
\label{sec:meta-language}

The meta-language employed by our logical framework manipulates labeled property graphs (LPGs), with deduction rules and proofs being encoded as graph transformations and LPGs, respectively. %
LPG is probably the most popular graph model used in practice, being used by NoSQL graph databases such as Neo4j, Memgraph, and Redis Graph (another popular graph model is RDF, or resource description framework). %
While graphs seen in algorithms and data structures courses are usually a simple pair of sets for vertices and edges (which are seen as (un-)ordered pairs of vertices), labeled property graphs let vertices and edges carry extra information that is useful in practice even if extraneous for most mathematical modeling.%
\footnote{Indeed, the extra information encoded in properties simplifies the reification of deductive rules in our LF and makes them more concise to express.} %

We give here the abstract definition of the labeled property graph model following \cite[§2.1]{besta2019demystifying}, but \cite[§2]{angles2017foundations} gives similar definitions. %
\cite[§4]{sinavier2019algebraic} gives a category theory-inspired definition similar in spirit to the one given here. %

The LPG model is an extension of the simple graph model mentioned earlier that adds metadata (labels and properties) to vertices and edges. %
Properties are key-value pairs, while labels are simple scalar values. %
A labeled property graph \(G\) is a tuple %
\begin{displaymath}
  (V, E, L, l_{V}, l_{E}, K, W, p_{V}, p_{E})
\end{displaymath}
where: %
\begin{itemize}
\item \(L\) is the set of labels; %
\item \(l_{V} : V \mapsto{} \mathcal{P}(L)\) and \(l_{E}: E \mapsto \mathcal{P}(L)\) are the labeling functions for vertices and edges respectively (\(\mathcal{P}\) is the power-set function); 
\item  \(K\) is the set of property keys; %
\item \(W\) is the set of property values; %
\item a property \(p\) is a pair \((k, v)\), with \(k \in K\) and \(v \in W\); %
\item \(p_{V} : V \mapsto \mathcal{P}(K\times{}W)\) and \(p_{E} : E \mapsto\mathcal{P}(K\times{}W)\) are the functions that give the set of properties of a given node and edge, respectively. %
\end{itemize}

Intuitively speaking, the usual graph model is simply some points and lines connecting them (the lines might or might not have arrows determining a direction for the relationship). In contrast, the labeled property graph model adds labels (which are helpful for description and documentation) and properties (which allow the storage of data related to vertex/edge in the vertex/node itself). %

In concrete terms, the meta-language employed by our LF is the Cypher query language, whose core has been formalized by \cite{francis2018cypher}. %
We write deductive rules as Cypher queries which manipulate proof graphs, and even formulas are represented as graphs. %


Taking inspiration from Racket's language-oriented programming \cite{felleisen2018racket}, we use Cypher as a low-level language, on top of which we build more high-level domain-specific languages (e.g., a small specification language for formulas that can be used to describe deduction rules). %
Most users of \glf{} will never need to learn any of these languages, relying instead on user interfaces described on \cref{sec:ui}. %
Expert users looking to develop their deductive systems would primarily deal with the high-level domain-specific languages, with a few of them needing to learn and use Cypher itself for more complex tasks (or to build other high-level languages). %

\subsection{Expressivity}
\label{sec:expressivity} %

Like most query languages, Cypher is a Turing-incomplete query language. %
Combining \glf{}'s strategy-specification language (described below) with raw Cypher queries is probably sufficient for Turing-completeness, but we are more interested in the expressivity of our domain-specific languages (hence the absence of a formal proof of this conjecture). %
\glf{}'s mini-languages are much simpler to use and are what we expect users wanting to implement systems to use. %
In this section, we give not a complete characterization of their expressivity and semantics (which is left as future work, see \cref{sec:future}), but a description of what they can do. %
As a first bound of the expressivity of \glf{}'s meta-language, we can say that it may implement any propositional Natural Deduction system that respects the sub-formula property (see \cite[§2, def. 1-4]{haeusler14subformula}) using its sub-formula specification mini-language. %
Since any system fitting the above constraints can be embedded in propositional minimal implicational logic (see \cite[§2, Lemma 1]{haeusler14subformula}), we can use \glf{}'s tactic language to specify a single proof search procedure for any such system.\footnote{Since it is an embedding and not an isomorphism \glf{} may not produce direct proofs in the system, but only in minimal-implicational logic.} %
Formalizing and proving the correctness and completeness of tactics implemented in \glf{} is planned future work (see \cref{sec:future}). %
\jeffi{temos uma prova do mapeamento de MIMP em GLF?}
\jeffi{citaria o artigo do hermann que tem essa prova: How Many Times do We Need an Assumption to Prove a Tautology in Minimal Logic?}
\paragraph{Proof tactic specification language}
Semiautomatic proofs can be attempted in \glf{} according to a specification expressed by a small tactic language based on combinators. %
A tactic does not specify a single proof attempt, but a subset of the proof space to explore; %
according to the tactic specification, it backtracks (or not) until a proof is found or the possibilities are exhausted. %
\begin{itemize}
\item \texttt{Atomic}\((r, a)\) (where \(r\) is a deduction rule) turns a single deduction rule into a tactic that applies the input rule once or fails if the rule does not apply. %
The optional \(a\) argument allows one to enumerate the possible arguments to the rule, which are then tried one by one until one succeeds.\todo{This can be used to override a rule's normal argument enumerator and apply it only when stricter conditions apply (e.g., when no cycle would arise)} %
When backtracking, a successful rule application may be reverted, and the next argument tried as a new proof attempt. %
\item \texttt{Many}\((t)\) specifies a new tactic that applies the input tactic as many times as possible, without failing. %
\item \texttt{Try}\((t)\) specifies a new tactic that attempts to apply the input tactic once and never fails. %
\item \texttt{AndThen}\((t_{1}, t_{2})\) specifies a new tactic that applies \(t_{1}\) and then \(t_{2}\) (as long as \(t_{1}\) does not fail), else fails. %
\end{itemize} %
With these combinators, we can define other, non-primitive combinators: %
\begin{itemize}
\item \texttt{Some}\((t) \equiv \texttt{AndThen}(t, \texttt{Many}(t))\) specifies a new tactic that applies the input tactic as many times as possible, but fails if no applications were possible. %
\item \texttt{OrElse}\((t_{1}, t_{2}) \equiv \texttt{AndThen}(\texttt{Try}(t_{1}), t_{2})\) specifies a new tactic that attempts to apply \(t_{1}\), and if that fails, applies \(t_{2}\). %
\end{itemize}

\paragraph{Sub-formula specification language} %
This DSL leverages the subformula property\todo{cite again?} to help define legal rule premises and specify new branches created by rule application -- their goal formulas and which hypotheses they have introduced. %
This formula specification is done by referring to formulas relative to some frame formula, usually the pre-rule application goal formula.\todo{link to documentation} %

\begin{definition}[Relative formula reference]
  Let \(\mathbb{F}\) be the set of all formulas. %
  A relative formula reference is a function \(r : \mathbb{F} \to 2^{\mathbb{F}}\), i.e., a function from the set of formulas to the powerset of formulas. %
  The following atomic combinators are used to construct a relative formula reference:
  \begin{itemize}
  \item \texttt{Identity}\((f) = \{f\}\), %
    refers to the frame formula itself; %

  \item \texttt{SuperOf}\((f) = \{g \mid \text{\(g\) is a superformula of \(f\)}\}\), %
    refers to any formula that is a superformula of the input reference; %

  \item \texttt{SubOf}\((f) = \{g \mid \text{\(g\) is subformula of \(f\)}\}\), %
    refers to any formula that is a subformula of the input reference;

  \end{itemize} %
  The following are compound combinators, which take a fixed number of relative formula references and return a new relative formula reference (the last argument is the input formula; think of the combinators as being curried). %
  \begin{itemize}
  \item \texttt{Both}\((r_{1}, r_{2}, f) = r_{1}(f) \cap r_{2}(f)\), %
    refers to the formulas that are referred by both \(r_{1}\) and \(r_{2}\); %

  \item \texttt{And}\((r_{1}, r_{2}, f) = \bigcup_{g \in r_{1}(f)} r_{2}(g)\), %
    refers to the formulas that are referred by \(r_{2}\) using the formulas referred by \(r_{1}\) as frames; %

  \item \texttt{That}\((r, f) = \{f \mid r(f))\}\), %
    refers to the frame formula provided that the set of formulas referred to by \(r\) is not empty.
  \end{itemize} %
\end{definition} %

In the concrete implementation of the DSL, some additions are made to simplify rule definition. %
The most notable additions are a way to specify the principal operator of a formula (in the case of all atomic operators), or what is the index of the operand being specified (in the case of \texttt{SuperOf} or \texttt{SubOf}). %
There is also an additional \texttt{Arg} combinator that ignores the frame formula and specifies the formula pertaining to one of the rule's premises instead. %

As an example of how to use the sub-formula specification language, we define the rule for implication introduction in a backward-directed Natural Deduction system (see \cref{fig:rule-example}). %
The proviso for implication introduction is that the goal formula is an implication; we can perform this check by specifying that the (otherwise unused) argument to the rule is the goal formula and that its principal operator is an implication. %
Because the frame formula by default is the goal formula, we only need the \texttt{Identity} combinator. %
This rule creates a single branch in the proof tree, whose goal formula is the second operator (the consequent) of the goal (implication) formula. %
It also introduces a hypothesis, which pertains to the first operator (the antecedent) of the last goal (implication) formula. %

\begin{figure}
  \centering
\begin{verbatim}
-- anything after two hyphens is a comment
-- let g be the frame (goal) formula
-- implication introduction rule:
Rule(-- a rule specifies its arguments as list (surrounded by square braces)
     -- argument named `implication' specifies g has the form x -> y
     args = ["implication" =: Identity(operator=->)],
     -- each branch has a goal formula and any number of hypotheses introduced
     branches = [NewBranch(goal = SubOf(operand=2), -- goal is y
                           newHypotheses = [SubOf(operand=1)])]) -- new hypothesis is x
\end{verbatim}
  \caption{Implication introduction rule in a backwards-directed Natural Deduction system}
  \label{fig:rule-example}
\end{figure}


\subsection{Proof representation}
\label{sec:proof-representation}

Every complete\footnote{The graph structure of incomplete proofs may vary from system to system and is for the time being an implementation detail.} proof stored in the backend has the same graph structure, composed of formula and deduction nodes. %
The graph structure presented here is inspired by that of \cite{quispe2014proof}. %

\begin{figure} %
  \centering %
  \begin{subfigure}{.45\textwidth} %
    \centering %
    \begin{displaymath} %
      \prftree[r]{\(\scriptstyle\implies\mathrm{I}\)}
      {\prftree[r]{\(\scriptstyle\implies\mathrm{I}\)} %
        {\prftree[r]{\(\scriptstyle\implies\mathrm{I}\)}
          {\prftree[r]{\(\scriptstyle\implies\mathrm{E}\)} %
            {{\prftree[r]{$\scriptstyle\land\mathrm{I}$}
                {\prfboundedassumption{A}} %
                {\prfboundedassumption{B}}
                {A \land B}}} %
            {\prfboundedassumption{A \land B \implies C}} %
            {C}} %
          {A \implies C}} %
        {B \implies (A \implies C)}} %
      {(A \land B \implies C) \implies (B \implies (A \implies C))}
    \end{displaymath}
    \label{fig:example-proof1}
  \end{subfigure}
  \begin{subfigure}{.5\textwidth}
    \centering %
    \includegraphics[width=0.9\linewidth]{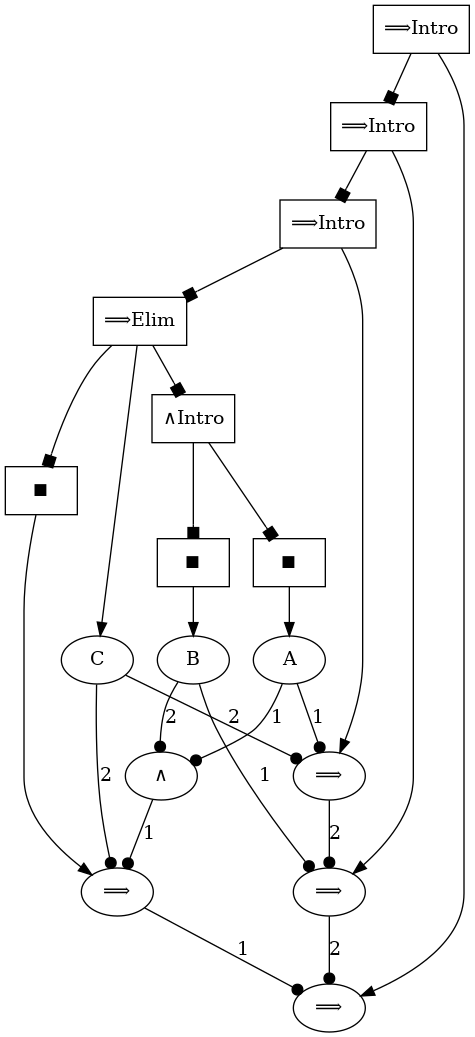}  %
    \label{fig:example-proof2}
  \end{subfigure} %
  \caption{Representation of the proof of \((A \land B \implies C) \implies (B \implies (A \implies C))\) (left) in the graph backend (right)} %
\label{fig:example-proof}
\end{figure}

\paragraph{Formula nodes}
Formulas are broken down into their constituent subformulas, which get represented as directed acyclic graphs. %
Atomic formulas are labeled with the atom they represent; compound formulas become nodes labeled by their principal operator. %
Additionally, compound formula nodes have incoming edges from their constituent formulas (therefore, atomic formulas are distinguished from compound ones by the absence of incoming edges from other formula nodes). %
In \cref{fig:example-proof} formula nodes are round-shaped, and the edges with dot-shaped arrowheads are the aforementioned formula-building edges. %
The node labeled \(\land\) represents the formula \(A \land B\), while the one labeled \(\implies\) represents \(A \land B \implies A\), as per the incoming dot-shaped arrowhead edges they have. %
It may help in understanding the representation of formulas to note that the representation is akin to a parse tree of the formulas in question, but with the sharing of common subformulas. %

\paragraph{Deduction nodes}
Deduction nodes may pertain to rule applications or to leaves in the proof tree, and always have an edge pointing to a formula node, indicating the formula derived in that step of the proof. %
Deduction nodes pertaining to a rule application are labeled by the rule's name and have outgoing edges to the roots of their child subproofs. %
The number of edges is determined by the number of premises a rule has; each such edge is labeled by the premise's role in the rule definition. %
In \cref{fig:example-proof} deduction nodes are box-shaped, and the edges with box-shaped arrowheads are the deduction edges. %
For instance, the box-shaped node labeled as \(\implies{}Elim\) pertains to the line in the proof where this rule is applied; the representation here is upside down in that we are going from the root to the leaves of the proof, so the node in question has two children (which are its premises in the regular proof). Because this node derives the formula \(C\), it has an arrow pointing to the formula node that represents this atom. %


\section{Related work}
\label{sec:discussion}


Although many logical frameworks have been developed, few have used graphs as a meta-language. %
\cite{breitner2016visual} is a graph-based generic theorem prover. %
It supports several logics, which can be specified using a simple text format. %
Although flexible, its design does not seem able to support sub-structural logics and maybe other more sophisticated deductive systems. %
It also does not offer the ability to customize its interface based on the logic being used, nor does it offer support for semiautomatic theorem proving (in the form of tactics or otherwise). %
It is meant as an educational tool, not as a formalization tool, and is not supposed to have good performance properties or handle large proofs (we have not ascertained whether it does or not, however). %

Porgy (presented in \cite{fernandez:hal-01251871}, among other works) is a graph rewriting system used to implement a visual modeling system and is not meant to be a logical framework, although it can conceivably be used as such. %
We could not personally test this system since the latest released version systematically crashed in our machine when following its tutorial. %
Porgy has a tactic specification language and thus supports semiautomatic use, but as far as we have read, it does not seem to support its use as a library and is thus confined to interactive use. %
It does not seem to support the programmatic specification of deductive systems, relying on its graphical interface for this task. %

\cite{schroeder2008graph} developed a system meant to be used as a logical framework from the start, complete with a language for specifying tactics. %
It is the first (to the best of our knowledge) graph-based logical framework but has not seen wide use since. %

Another system that is closely related to our work proposal is ELAN \cite{Borovansky1996elan}. %
It is a logical framework system implementation based on a many-sorted rewriting logic. %
Our proposal can be seen as a logical framework based on a graph rewriting system. %
Using graphs instead of an abstract rewriting system affords us the use of already implemented graph algorithms and makes the system easier to use and understand (since graphs are commonplace and intuitive). %

\section{Future work} %
\label{sec:future} %

In this paper, we have described a work-in-progress logical framework with a graph meta-language. %
Its foundations have been laid, but there are still many possible improvements to be made. %
A first step will be to implement more deductive systems besides the nine implemented so far; we intend to tackle first-order and non-structural logics first, then perhaps higher-order logics. %
Another point that goes hand-in-hand with the implementation of new logic is developing proof tactics for them. %
We expect such work to help us refine \glf{}'s tactic mini-language -- it currently demands the user write Cypher queries for more complex tasks like detecting cycles. %
We hope implementing tactics for diverse deductive systems will help us find the proper abstractions to encapsulate such tasks in the mini-language itself, freeing even advanced users from having to learn Cypher. %
Better ascertaining the expressivity of \glf{}'s mini-languages and formalizing such work is also planned. %

Another front of improvement is in the web-based user interface. %
It can certainly be refined from an aesthetic point-of-view, but most importantly, the user experience can be bettered by reducing the latency in user commands, which is especially bad when running it from a remote server instead of locally. %
The interface can also be made more intuitive to new users, and displaying formulas by default with infix notation instead of prefix is a simple step in this direction. %
We also plan on adding a fully generic user interface based on graphs that could work with any logic defined in \glf{}, instead of the specialized interfaces we currently have for backward Natural Deduction and Fitch-style systems. %

Finally, we plan future work on improving the time and space performance of \glf{}. %
This work ranges from database tuning (and comparison between the different possible backends for \glf{}) to the devising and implementation of algorithms for online and offline proof compression, which will be instrumental in helping the framework deal with huge proofs. %


\clearpage
\bibliographystyle{eptcs}
\bibliography{logic}

\end{document}